\documentclass[reprint,superscriptaddress,amsmath,amssymb,aps,prb,floatfix]{revtex4-2}
\usepackage{graphicx}
\usepackage{url}
\begin{document}

\title{Onset of the transitional flux-avalanche regime in bulk NbTi controlled by the thermal boundary conductance}
\author{I.~Abaloszewa}
\email{abali@ifpan.edu.pl}
\affiliation{Institute of Physics, Polish Academy of Sciences, Aleja Lotnik\'ow 32/46, PL-02668 Warsaw, Poland}
\author{V.~V.~Chabanenko}
\affiliation{O.~Galkin Donetsk Institute for Physics and Engineering, National Academy of Sciences of Ukraine, Kyiv, Ukraine}
\author{A.~Abaloszew}
\email{abala@ifpan.edu.pl}
\affiliation{Institute of Physics, Polish Academy of Sciences, Aleja Lotnik\'ow 32/46, PL-02668 Warsaw, Poland}
\date{\today}

\begin{abstract}
Thermomagnetic avalanches in type-II superconductors occur in two qualitatively different regimes, electromagnetically controlled in thin films and thermally limited in bulk samples, distinguished by the sign of the temperature derivative of the threshold field $H_\text{th}(T)$. The two regimes are separated by a critical thermal boundary conductance $h_c$, and a non-monotonic $H_\text{th}(T)$ has been predicted in the transitional region where the interface conductance $h$ approaches $h_c$. We approach this region in a bulk NbTi disk by raising the interface coupling above the pure-nonadecane baseline with a silver-filled interface layer. Whereas the pure interface gives a monotonically decreasing $H_\text{th}(T)$, the silver-filled interface produces a non-monotonic dependence not previously realized in a bulk superconductor: a temperature interval of positive slope, $dH_\text{th}/dT > 0$, terminating in a maximum at $T^* \approx 6.1$--$6.4$~K. The effect is reproduced for two independent silver-filled compositions; in a third, with the highest loading, the low-temperature decrease is absent altogether and $H_\text{th}(T)$ is flat up to the same $T^*$, the evolution expected for stronger coupling. The position of the maximum is set by the intrinsic properties of NbTi, independent of the silver content. The onset of the positive-slope interval coincides in temperature with a change of the avalanche morphology from narrow channeled fingers to broad fronts. The reversal of the sign of $dH_\text{th}/dT$ is a direct experimental signature of the onset of the transitional regime, in which heat removal during the instability becomes dynamically relevant.
\end{abstract}

\maketitle

\section{Introduction}

Thermomagnetic instabilities in type-II superconductors, abrupt avalanches of magnetic flux triggered by a positive feedback between Joule heating and critical current reduction, are one of the major limitations in applied superconductivity, where they can trigger quenches of superconducting magnets~\cite{SwartzBean,Wipf1967,Mints1981,Colauto2021,Xue2024}. They occur in many materials and geometries, from low-temperature superconductors such as Nb and NbTi to MgB$_2$ and high-temperature compounds, and have been characterized in detail by magneto-optical imaging of the evolving flux landscape~\cite{Colauto2021}. In thin films with strong thermal coupling to a substrate, the avalanches take the form of branching dendritic structures, propagate at velocities of tens of km/s on nanosecond timescales, and are controlled by electromagnetic diffusion~\cite{Vestgarden2018,Denisov2006,Bolz2003,Nulens2023}. In bulk superconductors, they are typically three orders of magnitude slower and thermally limited~\cite{Wertheimer1967,Vasiliev2006,Chabanenko2016,Chabanenko2022,Chabanenko2023SUST,Abaloszewa2026PRB}; recent combined experiments and coupled electromagnetic-thermal modeling of bulk MgB$_2$ magnets~\cite{Fracasso2024} have likewise identified the thermal exchange with the cold environment as the parameter that controls the onset of flux jumps~\cite{Fracasso2026}. These studies concern the onset of flux jumps, set by the thermal environment, within a single regime; the crossover region between the two regimes is not addressed.

The two regimes are distinguished by the sign of the temperature derivative of the threshold field for avalanche nucleation: $dH_\text{th}/dT < 0$ in the thermally limited regime~\cite{SwartzBean,Wipf1967,Abaloszewa2026PRB} and $dH_\text{th}/dT > 0$ in the electromagnetically controlled regime~\cite{Abaloszewa2023,Denisov2006}. In the linear-stability theory, the heat transfer coefficient at the superconductor-environment interface enters as the natural control parameter of the instability~\cite{Mints1981,Gurevich1987,Denisov2006}. For film samples, this coupling is provided by the substrate on which the film is grown, ensuring intimate thermal contact over the whole sample area; for bulk samples the thermal contact is made through a pressed or glued interface layer, and the resulting coupling is typically orders of magnitude weaker~\cite{Denisov2006}. In our recent work on a bulk NbTi disk~\cite{Abaloszewa2026PRB}, we established the thermally limited regime directly, through avalanche velocities of 15--25~m/s, millisecond development times, and a monotonically decreasing $H_\text{th}(T)$, and estimated, within the Mints--Rakhmanov (MR) framework~\cite{Mints1981}, the critical thermal boundary conductance $h_c$ separating the two regimes. For the pure nonadecane interface used there, $h \sim 10^3$~W/(m$^2\cdot$K) $\ll h_c \sim 10^5$~W/(m$^2\cdot$K). On this basis, Ref.~\cite{Abaloszewa2026PRB} predicted that intermediate coupling, $h \sim h_c$, should produce a non-monotonic $H_\text{th}(T)$, and proposed systematic variation of $h$ as the experimental test. The transitional region itself has remained experimentally untested.

In this work we implement this test. The measurement is made on the same bulk NbTi disk as in Ref.~\cite{Abaloszewa2026PRB}, so that the superconductor, its critical current density and the field window are unchanged and only the thermal interface differs: the coupling is raised above the pure-nonadecane baseline by adding silver powder to the nonadecane layer. We measure $H_\text{th}(T)$ for the pure interface and for three silver contents, compare the result with the quasi-static stability criterion, and follow the avalanche morphology over the same temperature range. The measurements provide evidence that the disk reaches the onset of the transitional regime.

\section{Sample and experimental setup}

The sample is a bulk NbTi disk (12~mm diameter, 0.1~mm thick, $T_c = 9.6$~K), the same as that used in Ref.~\cite{Abaloszewa2026PRB}. The threshold field was measured by magneto-optical imaging (MOI) using a Bi-doped ferrite garnet indicator film with an Al mirror layer, placed directly on the sample surface. The indicator, including its metallic mirror, was of the same type in all series and in Ref.~\cite{Abaloszewa2026PRB}, so that whatever thermal or electromagnetic role it plays is common to every measurement and cannot produce differences between compositions. The sample was cooled in zero field to the measurement temperature, and the external field was ramped at the maximum rate while the Hamamatsu ORCA~II camera recorded the flux distribution. The threshold field $H_\text{th}$ was identified as the applied field at which the first avalanche appeared in the image. The procedure for acquiring the $H_\text{th}(T)$ dependences was identical to that of Ref.~\cite{Abaloszewa2026PRB}.

The maximum field available at the sample is $\mu_0 H_a = 60$~mT, which is often below the threshold, in particular at the lowest temperatures. Whenever this was the case, the required excursion was obtained in two steps: the sample was cooled in a static negative field $\mu_0 H_\text{FC}$ (up to 30~mT in magnitude), and the field was then ramped to $+60$~mT, so that the total excursion $\mu_0 \Delta H = \mu_0 (H_a - H_\text{FC})$ exceeded the threshold. In the critical state, the screening-current distribution and the position of the flux front are determined by the change of the applied field relative to a state carrying no macroscopic currents~\cite{Bean1964,ClemSanchez1994}. Cooling in a uniform field establishes such a state, and a ramp by $\Delta H$ then reproduces the critical state of a zero-field-cooled sample subjected to the applied field $\Delta H$, up to a uniform offset of the local induction; cooling in a non-uniform field, by contrast, alters the screening capacity of the sample~\cite{Chaves2021}. The equivalence of the two protocols was verified directly: at a given temperature the threshold excursion, denoted $\mu_0 \Delta H_\text{th}$, does not depend on $H_\text{FC}$; for a series with $\mu_0 \Delta H_\text{th} = 50$~mT, avalanches appeared at $+25$~mT after cooling in $-25$~mT and at $+40$~mT after cooling in $-10$~mT. Since the amount of annihilated antiflux is proportional to $H_\text{FC}$ while the measured threshold excursion is not, this excludes an influence of the flux--antiflux annihilation that accompanies the ramp~\cite{Nulens2023,Chaves2024}; in both protocols the first avalanche nucleates at the sample edge. Points obtained with the two protocols are interleaved along each curve and show no systematic offset relative to one another.

The effective field evolution at the sample position was characterized in Ref.~\cite{Abaloszewa2026PRB}: eddy currents in the metallic components of the cryostat stretch the field establishment from the nominal 0.04~s to 0.1--0.4~s at cryogenic temperatures, limiting the maximum sweep rate at the sample to approximately 0.25~T/s at low temperatures. Two circumstances exclude an instrumental origin of the temperature structure reported below. First, in the interval 5--7~K the electrical resistance of the cryostat materials is on its residual plateau, so the instrumental field evolution is essentially identical across the temperature window in which the structure appears, and an instrumental artifact could not produce features at fixed temperatures. Second, for the pure nonadecane interface, $H_\text{th}(T)$ was shown to be independent of the sweep rate over more than three decades, from maximum-rate ramps down to $dH_a/dt = 0.1$~mT/s.

The dispersion of the data points visible in Fig.~\ref{fig:Hth} at each temperature is dominated by the intrinsic stochasticity of avalanche nucleation: each event develops from a different local flux configuration and at a different site along the sample edge, where nucleation occurs preferentially at weak spots~\cite{Silhanek2025,Brisbois2016}. A similar statistical dispersion of $H_\text{th}$ has been reported in Nb films~\cite{BlancoAlvarez}. A smaller contribution arises from the uncertainty in the instantaneous applied field, associated with the nonlinear character of the field ramp at low temperatures. Since the dominant source of spread is a physical property of the system rather than a measurement error, the individual events are shown explicitly. The point-to-point scatter along each curve serves as the estimate of this spread, and the non-monotonic variation of $H_\text{th}(T)$ exceeds it.

To implement the test proposed in Ref.~\cite{Abaloszewa2026PRB}, we modified the thermal interface by dispersing silver powder in the nonadecane matrix. Besides the pure nonadecane interface, three silver-filled compositions were studied, with silver contents of 10, 25, and 50~wt.\% (0.8, 2.5, and 7.0~vol.\%); with about 10~mg of nonadecane per layer and a weighing accuracy of 0.1--0.2~mg, the mass fractions are defined to within 2~wt.\%. All composites were prepared by weighing the components at the target ratios and mixing Ag powder (particle size of 1--10~$\mu$m) into nonadecane held at about 35\textdegree C, slightly above its melting point. The composite was applied to the cold finger while still molten and the sample was placed on top of it; upon cooling the layer solidified, providing both adhesion and a thermal interface. Volume fractions were calculated from the known densities of Ag ($\rho_\text{Ag} = 10.5$~g/cm$^3$) and solid nonadecane ($\rho_\text{n} \approx 0.79$~g/cm$^3$). Optical microscopy of the upper and lower surfaces of the solidified composite layers revealed no visible compositional gradient across the thickness.

Each composition corresponds to an independent mounting of the sample with a freshly prepared interface layer; the 50~wt.\% composition was measured twice within one mounting, at well-separated positions on the disk, with different indicator films and in separate temperature runs. The two runs agree within the scatter and are plotted with the same symbol in Fig.~\ref{fig:Hth}. The thickness of the layer and the placement of the indicator film cannot be reproduced exactly between mountings, so the absolute calibration of $H_\text{th}$ may differ from one series to another. The analysis below relies primarily on the shape of $H_\text{th}(T)$ within each series, and comparisons of absolute values between compositions are made with caution.

\section{Results}
\label{sec:results}

\begin{figure}
\includegraphics[width=\columnwidth]{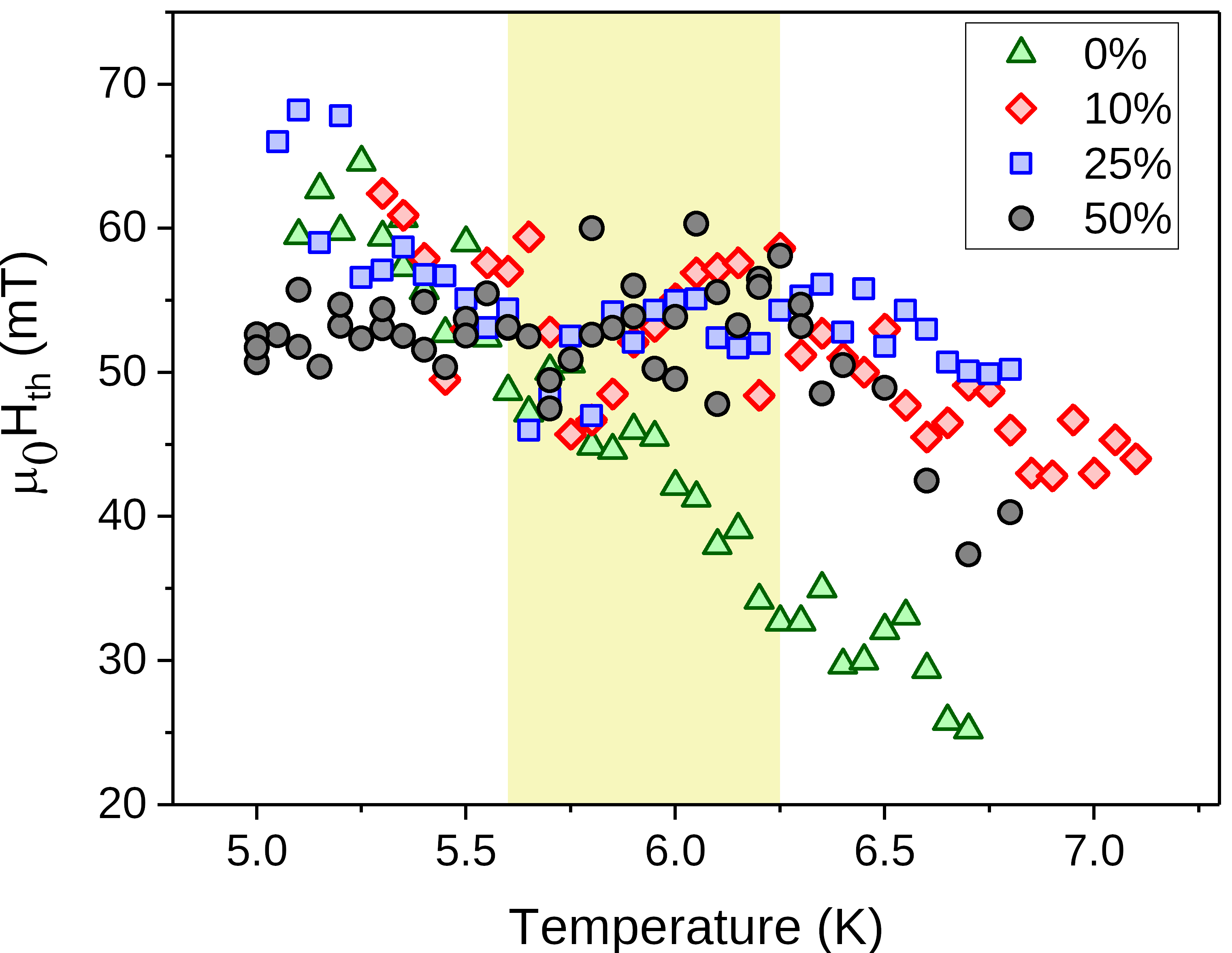}
\caption{Temperature dependence of the threshold field $\mu_0 H_\text{th}$ for avalanche nucleation in a bulk NbTi disk with pure nonadecane and three silver-filled nonadecane interfaces: 0\% (pure nonadecane), 10, 25, and 50~wt.\% Ag; the 50~wt.\% data combine two runs at different positions on the disk, with different indicator films, and are plotted with the same symbol. The shaded band marks the interval of positive slope, $dH_\text{th}/dT > 0$, that appears with the silver-filled interface.}
\label{fig:Hth}
\end{figure}

Figure~\ref{fig:Hth} presents $H_\text{th}(T)$ for the pure nonadecane interface and for three silver-filled compositions. The central result of this work is that the silver-filled interface changes $H_\text{th}(T)$ qualitatively: the monotonic decrease seen with the pure interface is interrupted, in every silver-filled composition, by an interval in which the threshold rises or stays flat.

For pure nonadecane, $H_\text{th}$ decreases monotonically from about 60~mT at 5.1~K to about 25~mT at 6.7~K. The dependence is close to linear with a slope of approximately $-24$~mT/K, and $dH_\text{th}/dT < 0$ within the scatter of the data, confirming the thermally limited regime established in Ref.~\cite{Abaloszewa2026PRB}.

The silver-filled interface changes the shape of $H_\text{th}(T)$ qualitatively. For 10 and 25~wt.\% Ag, the curve develops a non-monotonic structure: $H_\text{th}$ first falls with increasing temperature, passes through a minimum, then rises over a finite interval, $dH_\text{th}/dT > 0$, reaches a maximum at $T^*$, and falls again above it. For 10~wt.\% Ag the minimum lies near 5.8~K and the maximum near 6.1~K, and for 25~wt.\% Ag near 5.75 and 6.35~K. The rise from the minimum to the maximum is of order 10--20\%, but its precise value depends on the smoothing used to locate the extrema in the presence of point scatter, and we do not use it as a quantitative measure.

At 50~wt.\% Ag the change is more radical: $H_\text{th}$ no longer falls at low temperature. From 5.0 to 6.3~K the threshold is flat, the slope being $+1.8 \pm 1.3$~mT/K, to be compared with $-24$~mT/K for pure nonadecane; above 6.3~K it falls steeply, at $-33 \pm 7$~mT/K. The two runs of this composition, at different positions on the disk and with different indicator films, give the same flat branch, with mean levels of 52.9 and 52.7~mT.

Three features are robust. First, with the pure nonadecane interface $H_\text{th}$ decreases monotonically, while every silver-filled composition departs from this behavior: the fall is interrupted by an interval of positive slope at 10 and 25~wt.\%, and is absent altogether at 50~wt.\%. Second, the maximum, or the end of the flat branch at 50~wt.\%, occurs at the same temperature within the scatter for all three compositions, $T^* \approx 6.1$--$6.4$~K, indicating that its position is set by the intrinsic temperature dependences of the NbTi parameters ($j_c$, $C$, $\kappa$) rather than by the silver content. Third, on the high-temperature branch the silver-filled curves lie above the pure-nonadecane curve, by a factor of about 1.6 in the range 6.2--6.5~K.

\begin{figure}
\includegraphics[width=\columnwidth]{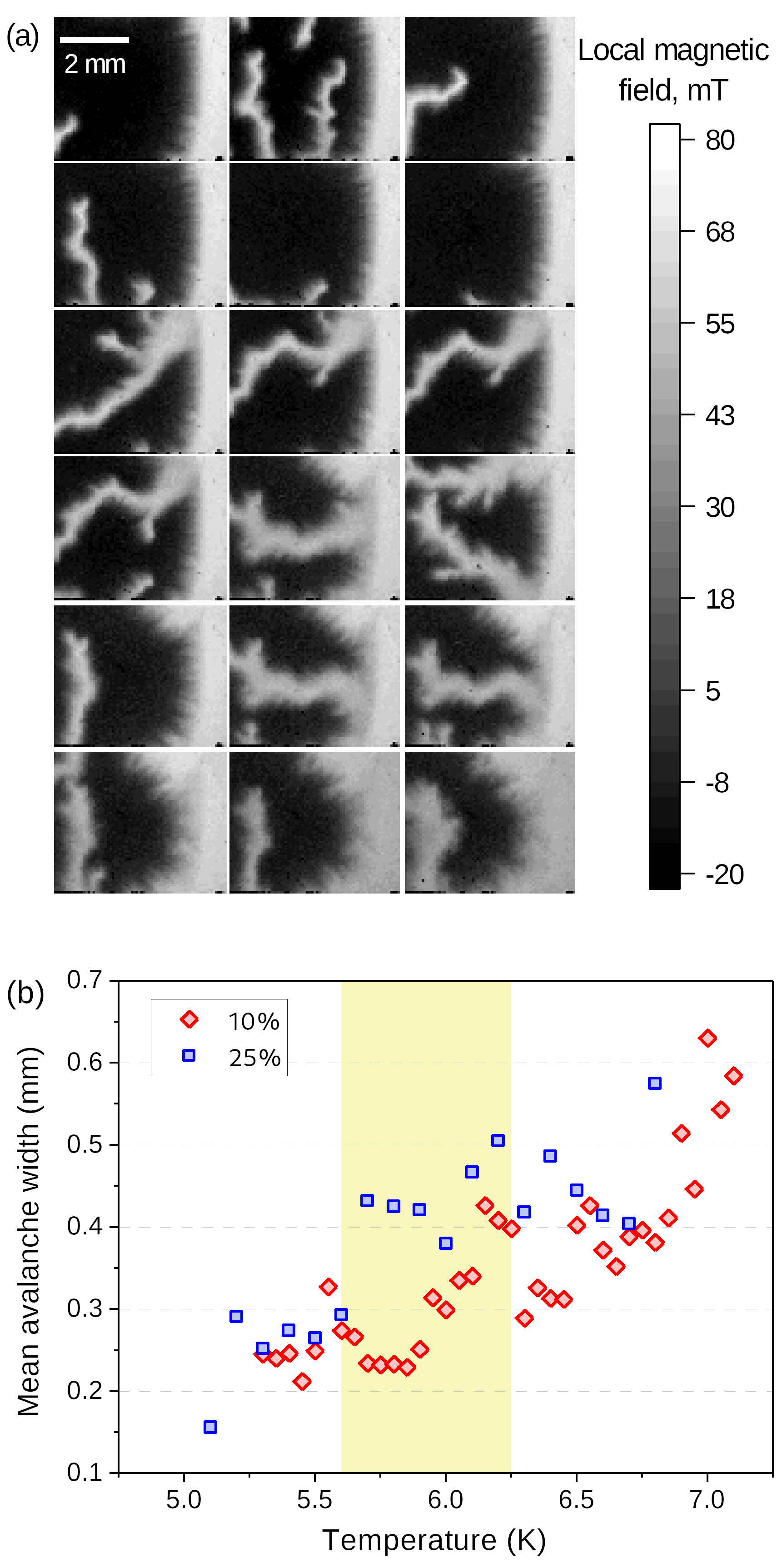}
\caption{(a) Magneto-optical images of the avalanche morphology in the 25~wt.\% Ag series, recorded from 5.1 to 6.8~K in 0.1~K steps (left to right, top to bottom); the right edge of each frame is the disk rim, the scale bar is 2~mm, and the grey scale gives the local magnetic field of the background-subtracted images. The morphology evolves from narrow channeled fingers at low temperature to broad fronts at high temperature. (b) Mean width of the penetrating structures versus temperature for the 10 and 25~wt.\% compositions; the shaded band marks the positive-slope window of Fig.~\ref{fig:Hth}.}
\label{fig:morph}
\end{figure}

The avalanche morphology evolves across the same temperature range, and the evolution is reproduced for the two compositions in which it was quantified, 10 and 25~wt.\% Ag. Magneto-optical images recorded in 0.1~K steps [Fig.~\ref{fig:morph}(a), 25~wt.\% Ag] show a change from narrow channeled fingers at low temperature to broad fronts at high temperature. Quantified as the mean width of the penetrating structures, this crossover appears for both compositions [Fig.~\ref{fig:morph}(b)]. The width is obtained by the standard local-thickness construction~\cite{Hildebrand1997}: the binarized structure is skeletonized and the local thickness, twice the Euclidean distance to the nearest background pixel, is averaged along the skeleton, the smooth flux band along the sample edge being excluded. This is a rotation-independent measure of the typical width of the structures rather than of their widest point. Below the positive-slope window the two independently mounted series give the same mean width, 0.26~mm. It rises to 0.44~mm for 25~wt.\% Ag and 0.31~mm for 10~wt.\% Ag within the window and to about 0.45~mm above it, an increase by a factor of about 1.7 for both compositions. At 25~wt.\% Ag the width steps up at 5.7~K; at 10~wt.\% Ag it rises more gradually from about 5.95~K. Both onsets fall within this interval. The width remains elevated above $T^*$, which is expected if it follows the approach to the temperature at which the instability disappears, above 6.8~K here, rather than the value of $H_\text{th}$ itself~\cite{Denisov2006}. 

The absolute values depend on the binarization threshold, but the growth factor is insensitive to the choice of averaging: taking the median, the widest inscribed circle, or excluding the branch points of the skeleton changes it by less than 15\%. The growth of the characteristic instability width upon approaching the stability boundary was identified in Ref.~\cite{Denisov2006} as a prediction of the linear theory that had not been tested quantitatively. A temperature dependence of the size and of the fractal dimension of the first dendrites has been documented in NbN films~\cite{Rudnev2005} and reproduced in simulations~\cite{Vestgarden2011}, but not the width itself. The present bulk data exhibit it directly, and its reproduction for two compositions ties the morphological crossover to the same temperature interval in which $dH_\text{th}/dT$ changes sign.

\section{Discussion}

The observations of Sec.~\ref{sec:results} raise two questions: how so small a filler content can change the thermal coupling at all, and why the resulting $H_\text{th}(T)$ is not reproduced by the quasi-static stability criterion. The two subsections below take them in turn.

\subsection{Thermal bridges and characteristic timescales}
\label{sec:bridges}

All three Ag concentrations are below the percolation threshold for random dispersions of spherical particles, $\phi_c \sim 15$--33~vol.\%~\cite{Forero2021}, which for silver in nonadecane corresponds to 70--87~wt.\%. In this dilute limit the effective thermal conductivity of the composite is given by the Maxwell--Garnett approximation~\cite{Maxwell1873,Nan1997}:
\begin{equation}
\kappa_\text{eff} = \kappa_m \cdot \frac{\kappa_p + 2\kappa_m + 2\phi(\kappa_p - \kappa_m)}{\kappa_p + 2\kappa_m - \phi(\kappa_p - \kappa_m)},
\label{eq:MG}
\end{equation}
where $\kappa_m \approx 0.10$~W/(m$\cdot$K) is the thermal conductivity of solid nonadecane~\cite{Rastgar2016} and $\kappa_p \approx 430$~W/(m$\cdot$K) is that of silver, both at room temperature. In the limit $\kappa_p \gg \kappa_m$ realized here, Eq.~(\ref{eq:MG}) reduces to $(1 + 2\phi)/(1 - \phi)$, so that the enhancement is fixed by the volume fraction alone and is insensitive to the values taken for the two conductivities. For the three volume fractions it is only about 3, 8, and 23\% over pure nonadecane, and the boundary conductance, estimated as $h \approx \kappa_\text{eff}/L$ for a layer thickness $L = 100$~$\mu$m, stays within $(1.0$--$1.2) \times 10^3$~W/(m$^2\cdot$K) for all three compositions, indistinguishable from the pure nonadecane value~\cite{Abaloszewa2026PRB}. The symmetric Bruggeman theory~\cite{Bruggeman1935,Nan1997} gives the same result in this dilute limit. This estimate is the reference for what follows: any effective-medium description of the composite predicts essentially no change of $h$ at these concentrations, while the measured $H_\text{th}(T)$ changes qualitatively (Sec.~\ref{sec:results}). The enhancement of the thermal coupling must originate from a mechanism beyond the homogeneous-medium picture, that is, from the spatial arrangement of the filler rather than from its average concentration.

For reference, the critical thermal boundary conductance of our disk is~\cite{Abaloszewa2026PRB}
\begin{equation}
h_c(T) = \frac{\rho_\text{ff}\, j_c^2(T)\, d}{T_c - T},
\label{eq:hc}
\end{equation}
which, evaluated with the parameters established there [$\rho_\text{ff} \approx 2.7 \times 10^{-9}$~$\Omega\cdot$m at $B \approx 0.06$~T, $j_c(6~\text{K}) \approx 1.2 \times 10^9$~A/m$^2$, Kramer scaling~\cite{Kramer1973} with $T_c = 9.6$~K], is $h_c \approx 1.8 \times 10^5$~W/(m$^2\cdot$K) at 5.0~K and $1.2 \times 10^5$~W/(m$^2\cdot$K) at 5.8~K, consistent with $h_c \approx 10^5$~W/(m$^2\cdot$K) at 6~K obtained there. These values set the scale against which the coupling of the composites is assessed below.

A plausible microscopic mechanism is provided by the discreteness of the filler. In the thin composite layer ($L \sim 100~\mu$m), chains or clusters of Ag particles can be in simultaneous mechanical contact with both the sample surface and the cold finger, acting as thermal bridges that provide direct metallic heat conduction across the interface, a contribution absent from any homogeneous-medium description. The thermal conductance of a single bridge of effective metallic cross section $A_\text{b} \approx (5~\mu\text{m})^2$ spanning a gap $L = 100~\mu$m is $g_\text{br} \approx \kappa_\text{Ag} A_\text{b} / L \approx 1 \times 10^{-4}$~W/K. If $N$ such bridges per unit area span the gap, the boundary conductance due to bridges is $h_\text{br} = N g_\text{br}$. The bridge density is not known independently, and we ask instead what density would be required. For the 25~wt.\% composition, $N \sim 10^2$~mm$^{-2}$ bridges would provide $h_\text{br} \sim 1 \times 10^4$~W/(m$^2\cdot$K), an order of magnitude above the bulk effective-medium estimate. The silver budget of such bridges is modest: for straight columns the volume fraction they occupy equals the area they cover, $\phi_\text{br} = N A_\text{b} \approx 0.25$~vol.\% of the layer, so this density requires only a small fraction of the silver present. Such bridges are sparse rather than dense: they cover 0.25\% of the interface area, with a mean separation of about 100~$\mu$m, twenty times their own size. The corresponding ratio is $h/h_c \approx 0.1$ at 5.8~K, an order of magnitude closer to the transition than for the pure nonadecane interface, where $h/h_c \approx 0.01$.

The silver content also bounds the coupling from above. Heat is carried by the silver, so that $h_\text{br} = \phi_\text{br} \kappa_\text{Ag}/L$, where $\phi_\text{br}$ is the fraction of the layer volume occupied by spanning bridges. Even if all the filler were arranged in straight columns across the layer, $h$ would reach only $0.3\,h_c$ at 10~wt.\% and $0.9\,h_c$ at 25~wt.\% with the room-temperature $\kappa_\text{Ag}$, and about twice these values if the low-temperature conductivity of the particles is taken at the limit set by boundary scattering over their own size. Any real chain is longer than $L$ and interrupted by contacts, so these figures are upper limits. Reaching $h_c$ would thus require essentially all the silver to stand in ideal columns; yet the structure develops, at the same $T^*$, already at 10~wt.\%. A bridge requires an unbroken sequence of particles in contact across the whole layer, so the number of such chains is expected to grow steeply and nonlinearly with the filler concentration, as in percolation. The coupling should therefore increase along the sequence 10, 25, 50~wt.\%. The data do not resolve the difference between the two lower loadings: the position of $T^*$ is set by the temperature dependences of the NbTi parameters and is insensitive to $h$, and we do not use the amplitude quantitatively. The onset of the crossover is inferred experimentally from the qualitative change of $H_\text{th}(T)$ between the pure and the silver-filled interface, and does not depend on the specific microscopic mechanism of the $h$ enhancement.

Within the bridge picture, the characteristic heat removal time is $\tau_h = C d / h \sim 10~\mu$s for the 25~wt.\% composition, between pure nonadecane with $\tau_h \sim 100$~$\mu$s~\cite{Abaloszewa2026PRB} and thin films on substrates with $\tau_h \sim 50$~ns~\cite{Vestgarden2012}. The measured avalanche development time is three orders of magnitude longer, $\tau_\text{av} \sim 0.5$--1~ms~\cite{Abaloszewa2026PRB}, so that heat is removed appreciably while a single avalanche develops, as required by the dynamical interpretation of Sec.~\ref{sec:breakdown}. Its ratio to the thermal runaway time, $\tau_h/\tau_\text{run} = h_c/h \sim 10$ with $\tau_\text{run} = C(T_c - T)/\rho_\text{ff} j_c^2 \sim 1~\mu$s at 6~K, is the same statement as $h/h_c \approx 0.1$ expressed in the time domain used for films, and carries the same order-of-magnitude uncertainty.

\subsection{Temperature dependence of the threshold field}
\label{sec:breakdown}

The appearance of a non-monotonic $H_\text{th}(T)$ upon enhancement of $h$ confirms the central prediction of Ref.~\cite{Abaloszewa2026PRB}, made within the MR stability framework~\cite{Mints1981}. Its specific form, however, requires closer examination. Within the linearized MR analysis, the threshold field for a slab of half-thickness $d$ with surface heat transfer coefficient $h$ can be written as an interpolation between the adiabatic and the dynamic limits,
\begin{equation}
H_\text{th}^2(T, h) = H_\text{ad}^2(T) \left(1 + \frac{h}{h_c(T)}\right),
\label{eq:MR}
\end{equation}
where $h_c(T)$ is given by Eq.~(\ref{eq:hc}) and $H_\text{ad}(T)$ is the adiabatic threshold~\cite{SwartzBean,Wipf1967}. Neither the temperature dependence of $H_\text{ad}$ nor that of $h$ is measured independently; both have to be assumed, and the shapes that Eq.~(\ref{eq:MR}) can produce depend on the dependences assumed (Appendix~\ref{app:scan}).
At $h = 0$, $H_\text{th} = H_\text{ad}$ and $dH_\text{th}/dT < 0$. As $h$ increases, the stabilizing factor $(1 + h/h_c)$ grows with temperature, because $h_c(T)$ decreases through $j_c(T)$. The competition between the decreasing $H_\text{ad}(T)$ and the increasing correction can in principle produce a single interior maximum of $H_\text{th}(T)$.

We therefore examined what shapes of $H_\text{th}(T)$ Eq.~(\ref{eq:MR}) can actually produce. At the coupling estimated for our composites, $h/h_c \approx 0.1$, the interpolation predicts no structure at all, that is, no interior extremum: for any adiabatic baseline consistent with the measured pure-nonadecane curve the result is monotonic, with a modulation of a few percent across the window (Appendix~\ref{app:scan}, where Eq.~(\ref{eq:MR}) is evaluated over a broad range of parameterizations of $H_\text{ad}$, $j_c$, and $h$). Interior maxima require $h/h_c \gtrsim 1$, an order of magnitude above the estimate, together with parameterizations of $H_\text{ad}$ and $h(T)$ that the data do not support. Even then, their position is set by the chosen parameterization rather than fixed by the theory. An interior minimum followed by a maximum, the shape actually observed, essentially never occurs (Appendix~\ref{app:scan}). The explicit linear-stability expressions for the uniform and the fingering branches~\cite{Mints1981,Rakhmanov2004,Denisov2006} are likewise monotonic in temperature and in the background electric field, so the threshold given by the lower of the two shares this restriction (Appendix~\ref{app:branches}, which collects the explicit expressions and their monotonicity). For 10 and 25~wt.\% Ag, the measured curves, in contrast, fall steeply below the positive-slope window, in parallel with the pure-nonadecane series, so that the observed maximum is necessarily preceded by an interior minimum, whatever its exact position and depth; and the observed rise of 10--20\% exceeds anything the static expression produces by an order of magnitude, a comparison that is insensitive to the smoothing used to locate the extrema. The known reentrant stability of superconducting films, in which the instability disappears above an upper field $H^*$~\cite{Rudnev2005} as well as above a threshold temperature, occurs as a function of the applied field, through the field dependence of $j_c$~\cite{Colauto2021}, and is distinct from the temperature structure reported here. The static description thus predicts, at the estimated coupling, the absence of the structure that is observed. The most natural interpretation is that the structure reflects heat removal that occurs dynamically, during the avalanche, and is absent from the quasi-static balance: this is the expected behavior in the transitional regime, where the relevant timescales become comparable and a full dynamical theory, of the type developed for thin films by numerical simulation~\cite{Vestgarden2012,Vestgarden2018}, is required to describe the amplitude. The non-monotonic structure is in this sense an experimental signature of the onset of the transitional regime.

Estimates of the background and local electric fields (Appendix~\ref{app:fields}, where the two fields are evaluated from the measured ramp rate and avalanche velocity) separate two stages of the instability and clarify how the interface enters. At the nucleation of the instability, the electric field induced by the measured field ramp is more than two orders of magnitude below the critical field for nonuniform development~\cite{Rakhmanov2004}. Nucleation proceeds, at all accessible sweep rates, in the uniform branch of the linear-stability analysis (Appendix~\ref{app:branches}), which is the branch into which surface cooling enters. This is why the interface conductance affects the threshold field at all. During the development, the local field generated by the propagating avalanche exceeds this critical field, and the event is prone to channeling into narrow structures. That a uniformly nucleated instability develops into a localized, dendritic pattern is established in dynamical simulations~\cite{VestgardenJLTP}. The threshold field is therefore set at the slow, cooling-sensitive stage, while the morphology reflects the fast, nonlinear development stage; the distinction between uniform and dendritic development has been confirmed experimentally in films~\cite{DenisovPRL2006}.

The essential feature of these data is not the minimum--maximum pair as such, but the appearance of a finite temperature interval in which $dH_\text{th}/dT > 0$. In the thermally limited regime, $H_\text{th}(T)$ decreases monotonically. The pure-nonadecane control excludes intrinsic origins of the reversal: the disk, its $j_c(T)$, and the field window are the same, and the response is monotonic; the mechanism must therefore reside in the interface. A metal-containing interface can stabilize the instability in two ways, by removing heat or by electromagnetic braking through eddy currents~\cite{Colauto2010}; braking requires a continuous conducting layer adjacent to the superconductor and is ineffective at the sub-percolation dilutions used here (Sec.~\ref{sec:bridges}), where isolated micrometer-scale particles support only negligible eddy currents. This leaves heat removal during the instability as the mechanism that raises the threshold and reverses the sign of the slope. The disk is electromagnetically thin ($d/\delta \approx 0.1$, the Brandt thin-disk regime~\cite{Brandt1994,Abaloszewa2026PRB}), so its flux dynamics are nonlocal as in thin films~\cite{Aranson2001}. What distinguishes it from the film case is the rate of heat removal, not the electrodynamics: raising $h$ drives the same nonlocal system from the thermally limited toward the cooling-controlled regime. The two-stage character of the event indicates how the reversal can arise without any change of the static conductance. The broad, nearly uniform events at higher temperatures develop on larger scales and longer effective times, and are expected to benefit from the interface coupling. The narrow, channeled events at lower temperatures are closer to adiabatic with respect to it. As the morphology crosses from channeled to broad over the same range (Sec.~\ref{sec:results}), the integrated heat removal becomes effective and $H_\text{th}(T)$ rises. 

The interior minimum that precedes the rise is shallow and varies between mountings, and we do not rely on it; a quantitative account of the shape, including whether a genuine interior minimum survives, is deferred to a dynamical theory of the transitional regime. With further increase of $h$, the rising branch should extend toward the lowest measured temperatures, leaving a single maximum at the same loading-independent $T^*$. This is what the 50~wt.\% composition shows: the low-temperature decrease is absent, $H_\text{th}(T)$ is flat up to $T^*$, and the fall sets in only above it (Sec.~\ref{sec:results}). The rising branch is absent for $h \ll h_c$, and the amplitude of the structure should acquire a dependence on the field sweep rate in the transitional regime, because the heat removed while the threshold is approached grows as the ramp is slowed, whereas in the thermally limited limit removal is irrelevant and no such dependence is expected; that the rate of field change can affect the onset of the instability has been demonstrated for films~\cite{Baziljevich2014}. This measurement is planned.

The compositions studied here are dilute, below the percolation threshold, where the filler acts on the thermal coupling alone; above about 70~wt.\% a percolating metal network would also screen electromagnetically and brake the flux dynamics~\cite{Colauto2010}, and the layer would no longer be a purely thermal element. Within this dilute regime the composites vary $h$ over a limited range and with limited control, so a systematic mapping of the $h$--$T$ phase diagram requires a more direct means of varying the interface conductance; this, together with the dynamical theory noted above, defines the direction of further work.

\section{Conclusion}

We have experimentally verified the central prediction of Ref.~\cite{Abaloszewa2026PRB}: raising the thermal boundary conductance of a bulk NbTi disk above the pure-nonadecane baseline makes $H_\text{th}(T)$ non-monotonic. The three silver-filled compositions trace the evolution expected as the coupling is increased. At 10 and 25~wt.\% Ag, an interval of rising threshold opens and closes at $T^*$, while at 50~wt.\% the low-temperature decrease has disappeared and the threshold is flat up to the same $T^*$. The position of $T^*$, 6.1--6.4~K, does not move with the silver content, and the shallow minimum that precedes the rise varies between mountings; the interval of rising threshold, rather than the minimum--maximum pair, is therefore the primary observation. At the estimated coupling the quasi-static Mints--Rakhmanov interpolation predicts no discernible structure (Sec.~\ref{sec:breakdown}); the existence, the shape, and the amplitude of the observed structure are most naturally attributed to heat removal acting dynamically during the avalanche. The structure is therefore a direct experimental signature of the onset of the transitional regime. It appears together with a crossover of the avalanche morphology from narrow channels to broad fronts, quantified here through the avalanche front width. A two-stage analysis of the instability, in which nucleation is uniform and sensitive to cooling while the fast development that follows tends to channel, offers a consistent qualitative mechanism, with a sweep-rate dependence of the amplitude as its testable consequence. Estimates based on discrete thermal bridges place the 25~wt.\% composition at $h/h_c \approx 0.1$, equivalently $\tau_h/\tau_\text{run} \sim 10$, consistent with the onset of the transition.

Together with the thermally limited regime of the pure interface ($h \ll h_c$) and the electromagnetically controlled regime of thin films ($h \gg h_c$)~\cite{Denisov2006,Abaloszewa2023}, the present data complete an experimental sequence of instability regimes spanned by a single control parameter. The shape of $H_\text{th}(T)$ thereby provides a non-invasive diagnostic of the thermal coupling of superconducting elements, refining the sign criterion proposed in Ref.~\cite{Abaloszewa2026PRB}, of relevance to trapped-field magnets~\cite{Chabanenko2026SUST} and to superconducting circuit components operating in magnetic fields~\cite{Nulens2023}. These results support the identification of the effective thermal boundary conductance as the control parameter governing the crossover between avalanche regimes in bulk type-II superconductors, and they define the target for a future dynamical theory of the transitional regime.

\section*{Data Availability}

The data that support the findings of this article are not publicly available upon publication because it is not technically feasible and/or the cost of preparing, depositing, and hosting the data would be prohibitive within the terms of this research project. The data are available from the authors upon reasonable request.

\appendix

\section{Numerical analysis of the quasi-static interpolation}
\label{app:scan}

To establish which shapes of $H_\text{th}(T)$ the interpolation of Eq.~(\ref{eq:MR}) can produce, we evaluated it numerically over a broad range of parameterizations. The adiabatic threshold was taken either as a power law, $H_\text{ad} \propto (1 - T/T_c)^p$ with $p$ from 0.5 to 1.5, or as a linear function matched to the measured slope of the pure nonadecane series. For the critical current density, two forms were used: Kramer scaling~\cite{Kramer1973}, $j_c \propto (1 - T/T_c)^{1.5}$ with the sample value $T_c = 9.6$~K~\cite{Abaloszewa2026PRB}, and the practical critical-surface fit of Bottura~\cite{Bottura2000} reduced to a fixed low field, which gives $j_c(T) \propto (1 - t^{n})^{\gamma - \alpha}$ with $t = T/T_{c0}$ and the parameters of the fit, $\alpha = 0.57$, $\gamma = 2.32$, $n = 1.7$, $T_{c0} = 9.2$~K. We note that the $b^{\alpha}$ factor of the fit makes it formally inapplicable in the limit of vanishing field, so only its temperature dependence at the fixed working field of about 50~mT is used here. The boundary conductance was allowed a power-law temperature dependence, $h \propto T^m$ with $m$ from 0 to 3, covering the range from a temperature-independent contact to a Kapitza-like behavior, and the ratio $h/h_c$ at 5.8~K was varied from 0.05 to 5.

The flux-flow resistivity entering $h_c(T) \propto \rho_\text{ff} j_c^2/(T_c - T)$ was taken either constant or as $\rho_\text{ff} \propto 1/B_{c2}(T)$ at the fixed working field.

\begin{figure}
\includegraphics[width=\columnwidth]{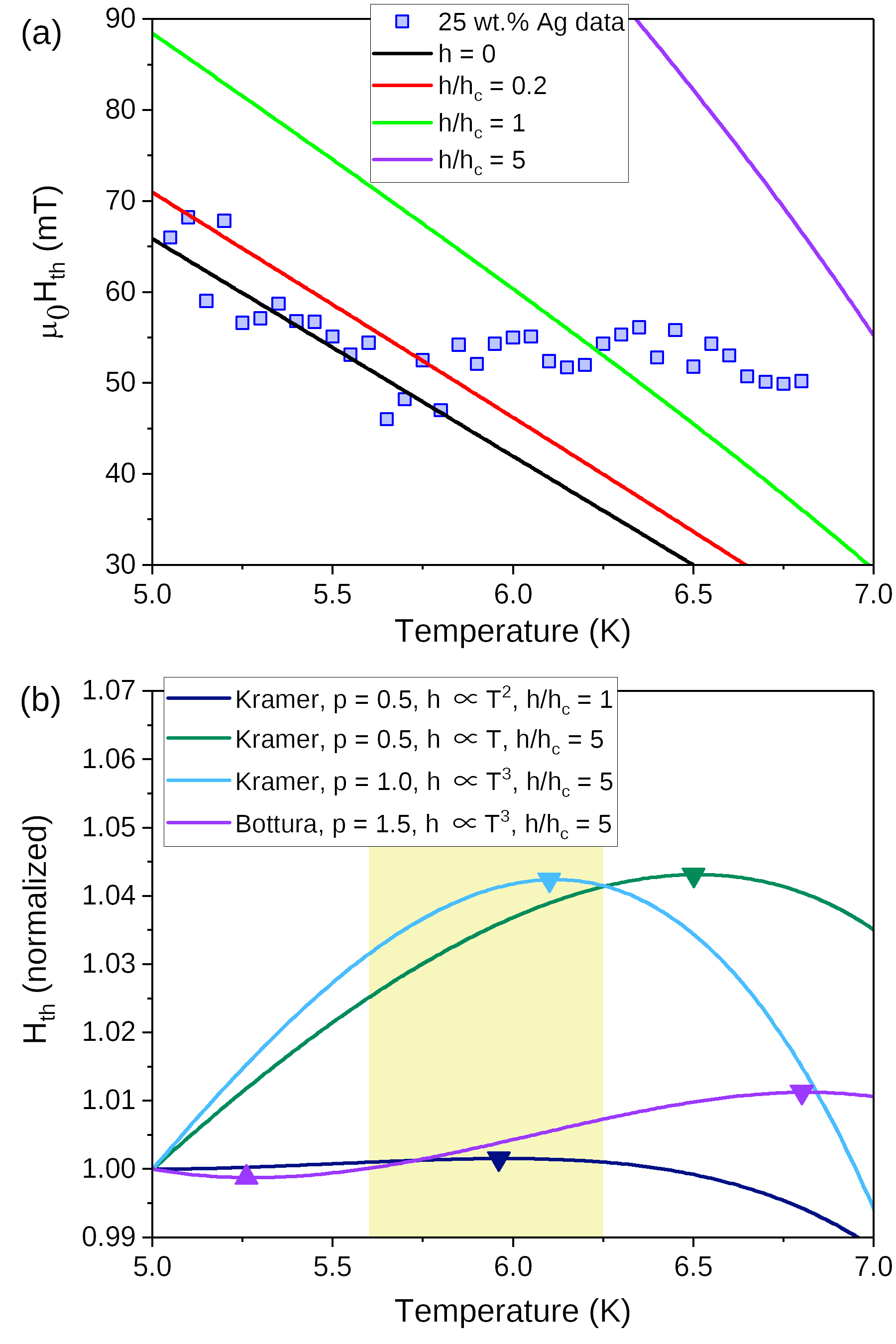}
\caption{Shapes admitted by the quasi-static interpolation, Eq.~(\ref{eq:MR}), in the measured window. (a) With the adiabatic threshold matched to the measured pure-nonadecane slope and $\rho_\text{ff} \propto 1/B_{c2}(T)$, the interpolation yields monotonically decreasing curves at all couplings up to $h/h_c = 5$, the values shown bracketing the coupling estimated for the composites, $h/h_c \approx 0.1$; the 25~wt.\% Ag data are shown for comparison. (b) The only parameterizations in the scan that produce an interior maximum (down triangles), all requiring $h/h_c \geq 1$, a strongly temperature-dependent $h$, and a weakly decreasing $H_\text{ad}$; the maxima scatter across the interval and only accidentally fall inside the positive-slope window of Fig.~\ref{fig:Hth} (shaded). The single parameterization producing a minimum (up triangle) followed by a maximum has a rise of 1.2\%.}
\label{fig:scan}
\end{figure}

Three properties hold across the entire grid (Fig.~\ref{fig:scan}). First, at the coupling estimated for our composites, $h/h_c \approx 0.1$, the interpolation predicts no structure at all: for every adiabatic baseline consistent with the measured pure-nonadecane curve the result is monotonic, and the correction factor modulates the baseline by only a few percent across the window [Fig.~\ref{fig:scan}(a)]. Second, interior maxima appear only for $h/h_c \gtrsim 1$, combined with a strongly temperature-dependent $h$ and a weakly decreasing $H_\text{ad}$, and their position is set by the chosen parameterization rather than fixed by the theory, scattering over 5.5--6.8~K across the grid [Fig.~\ref{fig:scan}(b)]. Third, an interior minimum followed by a maximum, the shape observed experimentally, occurs for exactly one corner of the grid ($p = 1.5$, $h \propto T^3$, $h/h_c = 5$), with a rise of 1.2\%, an order of magnitude below the observed 10--20\% and at a coupling more than an order of magnitude above our estimate.

The quasi-static interpolation thus acts as a null hypothesis: at the estimated coupling it predicts the absence of the structure that is observed.

\section{Linear-stability expressions for the two branches}
\label{app:branches}

The linear-stability theory of the thermomagnetic instability provides explicit threshold expressions for the uniform and the nonuniform (fingering) development. For a film of thickness $d$ and half-width $w$ on a substrate with heat transfer coefficient $h_0$, the uniform branch is~\cite{Denisov2006}
\begin{equation}
H_\text{uni} = H_\text{ad} \left(1 - \frac{2 T^{\,*} h_0}{n\, d\, j_c\, E}\right)^{-1/2},
\label{eq:Huni}
\end{equation}
where $E$ is the background electric field, $n$ the creep exponent, and $1/T^{\,*} = -\,d \ln j_c / dT$; the analogous expression for a slab follows from the dynamic criterion of Ref.~\cite{Mints1981}, and flux jumping in the thin-film geometry is treated in Ref.~\cite{MintsBrandt1996}. To first order in the cooling correction, Eq.~(\ref{eq:Huni}) reduces to the form of the interpolation, Eq.~(\ref{eq:MR}), with the background field evaluated at the flux-flow scale; the two expressions differ beyond first order, Eq.~(\ref{eq:Huni}) diverging at finite $h_0$ (full dynamic stabilization) while Eq.~(\ref{eq:MR}) remains finite, and Eq.~(\ref{eq:Huni}) retains $E$ as an explicit variable, which underlies the two-branch competition used in the main text. The correction decreases monotonically with $E$, so that within this branch the flux-jump field decreases monotonically with the electric field~\cite{Rakhmanov2004}. The fingering branch for a slab is~\cite{Rakhmanov2004}
\begin{equation}
H_\text{fing} = \frac{\pi}{2} \left(\frac{\kappa\, T^{\,*}\, j_c}{E}\right)^{1/2},
\label{eq:Hfing}
\end{equation}
with the corresponding film expression given in Ref.~\cite{Denisov2006}. The two expressions refer to different geometries, a film for Eq.~(\ref{eq:Huni}) and a slab for Eq.~(\ref{eq:Hfing}). For the present sample the slab expressions are the relevant ones: the film analysis of Ref.~\cite{Denisov2006} assumes a thickness below the London penetration depth, whereas our disk is three orders of magnitude thicker, and its geometric thinness, $d \ll R$, enters through the Brandt field distribution~\cite{Brandt1994} rather than through the stability expressions. The argument below uses only the monotonicity of the two branches, which holds in either geometry. Through the smooth temperature dependences of $j_c$, $C$, and $\kappa$, both branches are monotonic functions of temperature in the interval of interest, and the instability threshold, given by the lower of the two, can therefore exhibit at most a single interior extremum as a function of temperature.

\section{Electric fields and the two stages of the instability}
\label{app:fields}

The character of the instability development is controlled by the dimensionless ratio of thermal to magnetic diffusion, $\tau = \mu_0 \sigma \kappa / C$, where $\sigma$ is the differential conductivity at the working point~\cite{Rakhmanov2004}. The border between uniform and nonuniform development lies at $\tau = 1/n$~\cite{Rakhmanov2004}, with $n$ the creep exponent, which in dimensional form means that nonuniform development requires a background electric field exceeding
\begin{equation}
E_c \simeq \frac{\mu_0\, \kappa\, j_c}{C} \approx 0.05~\text{V/m},
\label{eq:Ec}
\end{equation}
where the numerical value corresponds to the parameters of Ref.~\cite{Abaloszewa2026PRB}, $\kappa \approx 0.03$~W/(m$\cdot$K) for NbTi, $C \approx 10^3$~J/(m$^3\cdot$K), and $j_c \approx 1.2 \times 10^9$~A/m$^2$, and is consistent in order of magnitude with the numerical example of Ref.~\cite{Rakhmanov2004}.

At nucleation, the background electric field is set by the field ramp. The flux penetration depth at threshold follows from the critical-state estimate for a thin disk of radius $R$~\cite{ClemSanchez1994},
\begin{equation}
l = R \left[1 - \frac{1}{\cosh(H_a/H_d)}\right],
\qquad
H_d = \frac{d\, j_c}{2},
\label{eq:lpen}
\end{equation}
which for $\mu_0 H_d \approx 75$~mT and $H_a \approx H_\text{th} \approx 50$~mT gives $l \approx 1$~mm. With the measured effective ramp rate at the sample, $\mu_0 \dot{H} \approx 0.1$--0.25~T/s~\cite{Abaloszewa2026PRB}, the background field is
\begin{equation}
E_\text{bg} = \mu_0 \dot{H}\, l \approx (1\text{--}2.5) \times 10^{-4}~\text{V/m},
\label{eq:Ebg}
\end{equation}
more than two orders of magnitude below $E_c$. The corresponding diffusion ratio at nucleation, with a creep exponent $n \approx 50$--80, typical for NbTi in this temperature range, is $\tau = E_c/(n E_\text{bg}) \approx 2.5$--10, more than two orders of magnitude above the border value $1/n$; for the slow ramps of Ref.~\cite{Abaloszewa2026PRB}, $dH_a/dt = 0.1$~mT/s, one obtains $\tau \sim 10^4$. Nucleation therefore proceeds, at all accessible sweep rates, in the uniform, cooling-sensitive branch.

During the development, the local electric field is generated by the avalanche propagation itself,
\begin{equation}
E_\text{loc} = v B \approx 1~\text{V/m},
\label{eq:Eloc}
\end{equation}
using the measured velocities $v = 15$--25~m/s and $B \approx 50$~mT~\cite{Abaloszewa2026PRB}. This exceeds $E_c$, and the corresponding ratio, $\tau = E_c/E_\text{loc} \approx 0.05$ with $n = 1$ in the flux-flow regime, places the developing avalanche in the regime prone to channeling into narrow structures. The threshold of the instability is thus determined at the slow, cooling-sensitive stage, while the morphology of the event reflects the fast, nonlinear development stage.

\end{document}